\documentclass[noshowpacs,noshowkeys, groupedaddress,twocolumn, floatfix,aps, prl,reprint]{revtex4-1}
\usepackage[english]{babel}
\usepackage{fontenc}
\usepackage{graphicx}
\usepackage{amsmath}
\usepackage{epstopdf}
\usepackage{amssymb}
\usepackage{amsbsy}
\usepackage{amscd}
\usepackage{color}
\usepackage{bbm}
\usepackage{braket}
\usepackage{bm}
\usepackage{siunitx}
\usepackage[normalem]{ulem} 
\usepackage{verbatim}
\usepackage{url}
\usepackage[verbose,hypertexnames=false,bookmarksopenlevel=1,filecolor=blue,
linkcolor=blue,citecolor=blue,urlcolor=blue,pdfstartview=FitH,bookmarksopen,bookmarksnumbered,
colorlinks,plainpages=false,linktocpage]{hyperref}
\DeclareSymbolFont{mathbold}{OML}{cmm}{b}{it}
\DeclareMathSymbol{\bsigma}{\mathord}{mathbold}{27}
\begin{document}
\renewcommand{\figurename}{Fig.}
\title{Control of Spin Helix Symmetry in Semiconductor Quantum Wells by Crystal Orientation}
\author{Michael Kammermeier}
\email{michael1.kammermeier@ur.de}
\author{Paul Wenk}
\author{John Schliemann}
\affiliation{Institute for Theoretical
  Physics, University of Regensburg, 93040 Regensburg, Germany}
\date{\today }
\begin{abstract}
  We investigate the possibility of spin-preserving symmetries due to
  the interplay of Rashba and Dresselhaus spin-orbit coupling in $n$-doped
  zinc-blende semiconductor quantum wells of general crystal
  orientation.  It is shown that a conserved spin operator can be
  realized if and only if at least two growth-direction Miller indices agree in
  modulus. The according spin-orbit field has in general both
  in-plane and out-of-plane components and is always perpendicular to
   the shift vector of the corresponding persistent
  spin helix. We also analyze higher-order effects arising
  from the Dresselhaus term, and the impact of our results on weak
  (anti)localization corrections.
\end{abstract}
\maketitle
%
%
\allowdisplaybreaks
\paragraph{Introduction.} Extending the spin lifetime is  essential for the facilitation of spintronic 
devices~\cite{Awschalom2007}.
In semiconductors, owing to spin-orbit coupling (SOC) and impurity scattering, 
spin-polarized electrons or holes are subject to relaxation of their spin.
However, as a consequence of the interplay of Rashba and Dresselhaus SOC and, 
if necessary, strain or curvature effects, particular parameter configurations 
can be found~\cite{Schliemann2003, Bernevig2006,Trushin2007, Sacksteder2014,Dollinger2014,Wenk2016}.
These cases give rise to spin-preserving symmetries that remain  intact in the presence of
spin-independent disorder.
Their existence has been unambiguously confirmed in numerous experiments by 
means of optical and transport measurements~\cite{Koralek2009,Kunihashi2009,Faniel2011,WalserNat2012,Kohda2012,Ishihara2014,Yoshizumi2016}. 
The latter exploit the impact of SOC on weak (anti)localization.
A recent review summarizes the key developments in both theory and 
experiment~\cite{Schliemann2016}.

For a two-dimensional electron gas (2DEG), with SOC being linear
in wave vector $\mathbf{k}$, such a scenario leads to a SU(2) symmetry of the 
Hamiltonian $\mathcal{H}$, yielding persistent solutions for the spin 
diffusion equation with spin densities that are either homogeneous or 
helical in coordinate space.
This special symmetry is characterized by circular Fermi contours 
$\varepsilon_{\pm}$, shifted by a constant wave vector $\mathbf{Q}$, i.e., 
$\varepsilon_{-}(\mathbf{k})=\varepsilon_{+}(\mathbf{k}+\mathbf{Q})$ 
and by a spin-orbit field (SOF) which is collinear in $k$ space.
As a consequence, the spin of electrons traversing the system
undergoes a  well-defined rotation about the constant direction of 
the SOF,
which is independent of the propagated path, but solely determined by the 
initial and final position~\cite{Schliemann2003}, a phenomenon known by now
as the \textit{persistent spin helix}~\cite{Bernevig2006}.

The well-established cases of the above scenario are restricted to quantum wells grown along 
[001], [110], or [111] direction, where the SOF 
is either purely in-plane, purely out-of-plane, 
or vanishes, respectively~\cite{Zutic2004a,Schliemann2016}.
However, as we shall see in this Letter, also low-symmetry growth 
directions allow for such situations, and the  orientation of the SOF
with respect to the surface normal~$\mathbf{n}$ can in principle be designed 
arbitrarily.
This opens a wide range of possibilities for engineering spintronic devices.

\paragraph{Model Hamiltonian.} We consider a 2DEG whose crystal orientation 
is defined by an arbitrary normal unit vector $\mathbf{n}=(n_x,n_y,n_z)$
with the underlying basis vectors $\boldsymbol{\hat{x}}$, 
$\boldsymbol{\hat{y}}$, and $\boldsymbol{\hat{z}}$ pointing along the
crystal axes [100], [010], and [001], respectively.
The Hamiltonian describing the lowest conduction subband in an infinite quantum 
square well is given by
\begin{align}
\mathcal{H}={}&\frac{\hbar^2k^2}{2 m}+\boldsymbol{\Omega}\cdot\boldsymbol\sigma ,
\label{eq:hamiltonian}
\end{align}
where $m$ is the effective electron mass and $\boldsymbol\sigma$  
denotes the vector of Pauli matrices.
The effects due to Rashba (R)  and Dresselhaus (D) SOC
are comprised in the SOF 
$\boldsymbol{\Omega}=\boldsymbol{\Omega}_\text{R}+\boldsymbol{\Omega}_\text{D}^{(1)}+\boldsymbol{\Omega}_\text{D}^{(3)}$ with the dominant contributions
\begin{align}
\boldsymbol{\Omega}_\text{R}={}&\alpha\left(\mathbf{k}\times\mathbf{n}\right),\quad
\boldsymbol{\Omega}_\text{D}^{(1)}={}\beta^{(1)}\,\boldsymbol{\kappa},\label{eq:SOF}
\end{align}
where $\kappa_x=2n_x(n_yk_y-n_zk_z)+k_x(n_y^2-n_z^2)$ and analogous for the other components by cyclic index permutation~\cite{Dyakonov1986,Zutic2004a}.
In this formulation, the electron wave vector $\mathbf{k}$ 
is constrained by $\mathbf{k}\cdot\mathbf{n}=0$.
The field coefficients are given by $\alpha=\gamma_\text{R}\,\mathcal{E}_0$ and $\beta^{(1)}=\gamma_\text{D}\left[(\pi/a)^2-k^2/4\right]$.
Hereby, the Rashba SOC strength $\alpha$ is characterized by an electric field $\boldsymbol{\mathcal{E}}=\mathcal{E}_0\mathbf{n}$ as a result of a potential gradient in growth direction $\mathbf{n}$ of the quantum well.
In contrast, the Dresselhaus parameter $\beta^{(1)}$ strongly depends on the quantum well width $a$.
Additionally, both SOC coefficients are scaled by a material and confinement specific parameter~$\gamma_i$.
In the definition of $\beta^{(1)}$, the result of D'yakonov \textit{et al.}, Ref.~\cite{Dyakonov1986}, is extended by including the effect of $k$-cubic Dresselhaus terms, focusing only on the lowest angular harmonics in $k$. 
The $k$-cubic terms reduce the Dresselhaus SOC strength $\beta^{(1)}$ by a factor 
that depends on the wave vector $k$ which was already observed in Refs.~\cite{Iordanskii1994,Kettemann2007a} for [001] 2DEGs. 
The impact of $k$-cubic Dresselhaus terms w.r.t. higher angular harmonics is described by the field $\boldsymbol{\Omega}_\text{D}^{(3)}$.
Commonly, these terms constitute an obstacle for the realization of SU(2) symmetry.
We observe that only  the [111] and [110] growth directions allow us to construct a collinear SOF despite of the presence of $\boldsymbol{\Omega}_\text{D}^{(3)}$.
Yet, since the contribution $\boldsymbol{\Omega}_\text{D}^{(3)}$ is 
usually very small, it will be neglected hereafter.
It is discussed in more detail in the Supplemental Material.
\begin{figure}
\includegraphics[width= .9\columnwidth]{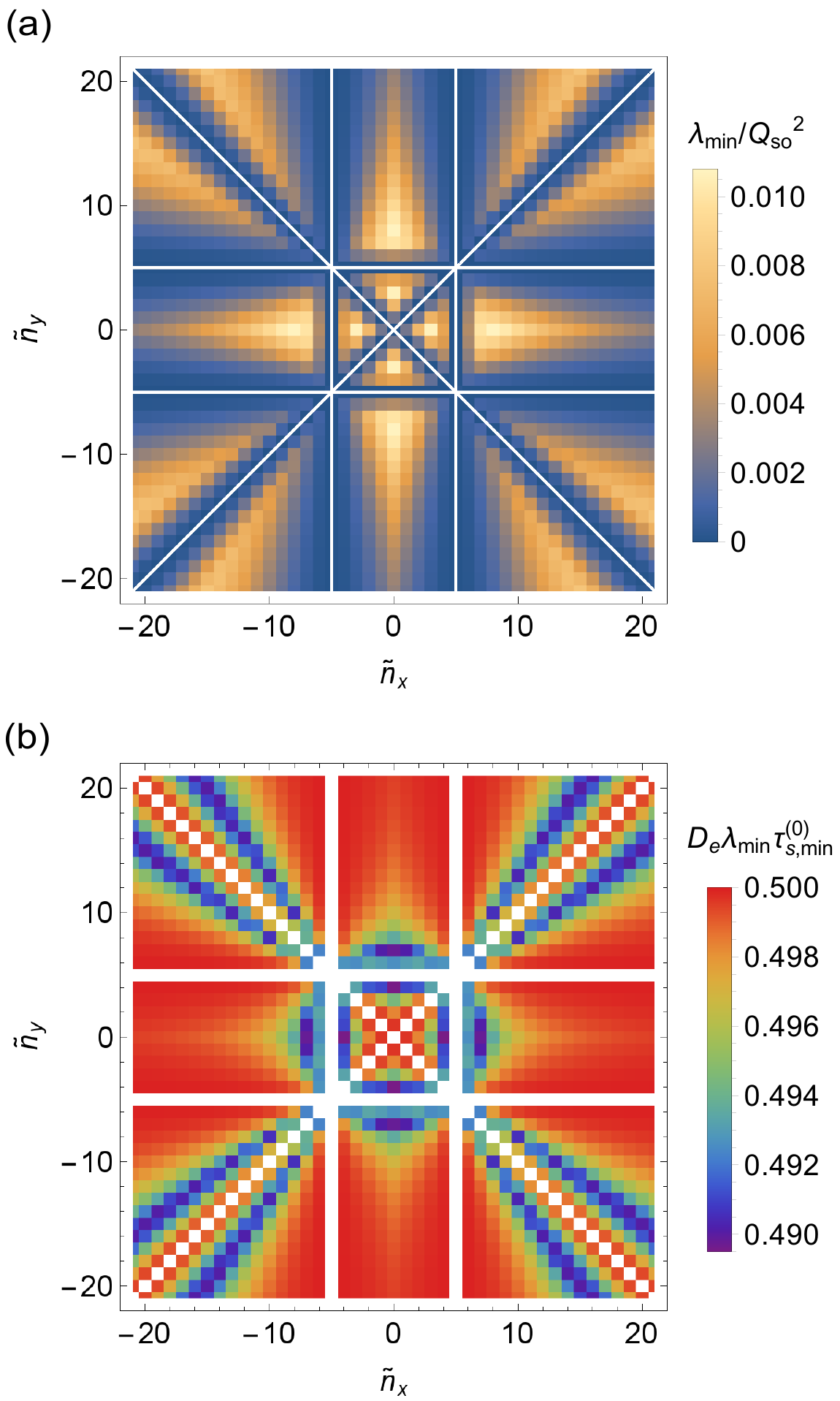}\\
  \caption{(Color online) (a) Global minimum $\lambda_\text{min}$ (in terms of $Q_\text{so}=4m\beta^{(1)}/\hbar^2$) of the spectrum of the spin diffusion operator $\Lambda_\text{sd}$ for the optimal ratio of Rashba and Dresselhaus coefficients $\alpha/\beta^{(1)}$ for different growth directions $\left[\tilde{n}_x,\tilde{n}_y,5\right]$ ($\tilde{n}_x,\tilde{n}_y\in \mathbb{Z}$). The white lines emphasize the vanishing minima.
  (b) Ratio between the global minimum $\lambda_\text{min}$  and the minimum $1/(D_\text{e} \hat{\tau}_\text{s,min}^{(0)})$ found by considering the spectrum at $\mathbf{q}=0$ solely, 
which corresponds to the D'yakonov Perel' spin-relaxation tensor, Eq.~(\ref{eq:SRtensor}). Along the white lines, both minima vanish exactly due to the SU(2) symmetry.}
  \label{plot:minima}
\end{figure}
\paragraph{Spin diffusion equation.} To gather information about the spin relaxation, we study the impact of SOC on the spin diffusion equation for weak SOC and disorder in the regime of zero temperature.
Selecting the Fourier representation with small wave vectors $\mathbf{k}$, $\mathbf{q}$, and frequencies $\omega$, leads to the equation for the spin density~$\mathbf{s}(\mathbf{q},\omega)$  \cite{Malshukov2005,Schwab2006,Wenk2010}:
\begin{align}
0={}&\left(D_\text{e} q^2-i\omega+1/\hat{\tau}^{(0)}_\text{s}\right)\mathbf{s}+\frac{4i\tau_\text{e}}{m}\braket{(\mathbf{q}\cdot\mathbf{k})\,\mathbf{\Omega}}\times\mathbf{s}.
\label{eq:spindiff}
\end{align}
Here, $\tau_\text{e}$ denotes the mean elastic scattering time, $D_\text{e}=v_\text{F}^2\tau_\text{e}/2$ the 2D diffusion constant with the Fermi (F) velocity $v_\text{F}=\hbar k_\text{F}/m$.
The corresponding D'yakonov Perel' spin-relaxation tensor is given by~\cite{Dyakonov1971,pikustitkov}
\begin{align}
\left( 1/\hat{\tau}_\text{s}^{(0)}\right)_{ij} ={}&
  4\tau_\text{e} /\hbar^2(\braket{\boldsymbol{\Omega}^2}\delta_{ij}-\braket{\Omega_i\Omega_j}).
  \label{eq:SRtensor}
\end{align}
The averaging $\braket{\ldots}$ is performed over all in-plane directions of $\mathbf{k}$ using the relation $\braket{k_ik_j}=(k_\text{F}^2/2)(\delta_{ij}-n_in_j)$~\cite{Zutic2004a}.
It is practical to rewrite Eq.~(\ref{eq:spindiff}) by means of the spin-diffusion operator $\Lambda_\text{sd}(\mathbf{q})$, i.e., $0=[D_e \Lambda_\text{sd}(\mathbf{q})-i\omega]\,\mathbf{s}$.
Parameter configurations which yield a vanishing eigenvalue $\lambda$ 
of $\Lambda_\text{sd}$ at a specific $\mathbf{q}=\mathbf{q}_\text{min}$ lead to an infinite spin lifetime. 
Thereby, we distinguish two cases depending on  $\mathbf{q}_\text{min}$:
(i) for $\mathbf{q}_\text{min}=0$, the long-lived spin state does not precess in coordinate space, 
(ii) for $\mathbf{q}_\text{min}=\mathbf{Q}\neq 0$, a persistent spin helix is formed.

\paragraph{Conditions for persistent spin states.}
In Fig.~\ref{plot:minima}(a) we display the global minimum $\lambda_\text{min}$ of the spectrum of $\Lambda_\text{sd}$ in dependence of the 2DEG orientation.
It is determined by identifying individually for a 2DEG with the Miller indices $\left[\tilde{n}_x,\tilde{n}_y,5\right]$ ($\tilde{n}_x,\tilde{n}_y\in \mathbb{Z}$), the optimal ratio of $\alpha/\beta^{(1)}$.
Along the white lines,  $\lambda_\text{min}$ vanishes exactly.
This indicates that a vanishing eigenvalue $\lambda$ demands at least two equal indices $\vert n_i\vert$ of the normal vector $\mathbf{n}$. 
Rigorous analytical calculations confirm this supposition (see Supplemental Material).

Thus, without loss of generality we restrict our analysis to the first octant, i.e., $n_i>0$, and for simplification set $n_x=n_y\equiv \eta$ and $n_z=\sqrt{1-2\eta^2}$ due to normalization.
The relation to the polar angle $\theta$ w.r.t.  $[001]$ is given by $\eta=\sin(\theta)/\sqrt{2}$.
Hence, the growth direction is defined by a plane which comprises all commonly known cases that allow for spin-preserving symmetries, i.e., [001], [111], and [110].
For an arbitrary $\eta\in \left[0,1/\sqrt{2}\right]$ the Rashba and Dresselhaus coefficients need to fulfill the relation
\begin{align}
\alpha/\beta^{(1)}={}&\Gamma_0:=(1-9 \eta^2)\sqrt{1-2 \eta^2}.
\label{eq:ratio}
\end{align}
Inserting this particular condition in Eq.~(\ref{eq:hamiltonian}), we can rewrite the Hamiltonian in a form which reveals the SU(2) symmetry, that is,
\begin{align}
\mathcal{H}={}&\frac{\hbar^2}{2 m}\left[k^2+\left(\mathbf{k}\cdot\mathbf{Q}\right)\Sigma\right].
\label{eq:hamiltonian2}
\end{align}
The spin operator
\begin{align}
\Sigma={}&\left(\sigma_x+\sigma_y+\frac{3\eta\sqrt{1-2\eta^2}}{3\eta^2-1}\sigma_z\right)/N\equiv \mathbf{\boldsymbol\Pi} \cdot \boldsymbol\sigma,
\label{eq:sigma}
\end{align}
with the normalization constant $N=\sqrt{2-3\eta^2}/\vert 1-3\eta^2\vert$ is a conserved quantity, i.e., $\left[\mathcal{H},\Sigma\right]=0$. 
 The direction of the collinear SOF is determined by the vector~$\boldsymbol\Pi$.
 As a result, it is always perpendicular to the $[\overline{1}10]$ axis and, thus, also to
 the wave vector 
\begin{align}
\mathbf{Q}={}&\frac{Q_0}{\sqrt{2}}(-1,1,0),
\end{align}
with $Q_0={}\vert 1-3\eta^2\vert\sqrt{1-3\eta^2/2}\,Q_\text{so}$ and $Q_\text{so}=4 m\beta^{(1)}/\hbar^2$, which induces the shift of the Fermi contours and describes the spin precession of the propagating electrons.
The length $L_\text{s}:=2\pi/Q_0$ is denoted as spin precession length. 
It specifies the distance along $\mathbf{Q}$ that spin-polarized electrons  need to propagate until their spin has performed a full precession cycle.
The corresponding precession axis is given by the orientation of~$\boldsymbol\Pi$.
Note that an additional solution occurs for $\eta=0$, that is, $\alpha=-\beta^{(1)}$, which results in $\mathbf{Q}=Q_0(1,1,0)/\sqrt{2}$ and $\Sigma=(\sigma_x-\sigma_y)/\sqrt{2}$ for a [001] confined 2DEG. 
 
\begin{figure}
  \includegraphics[width= .9\columnwidth]{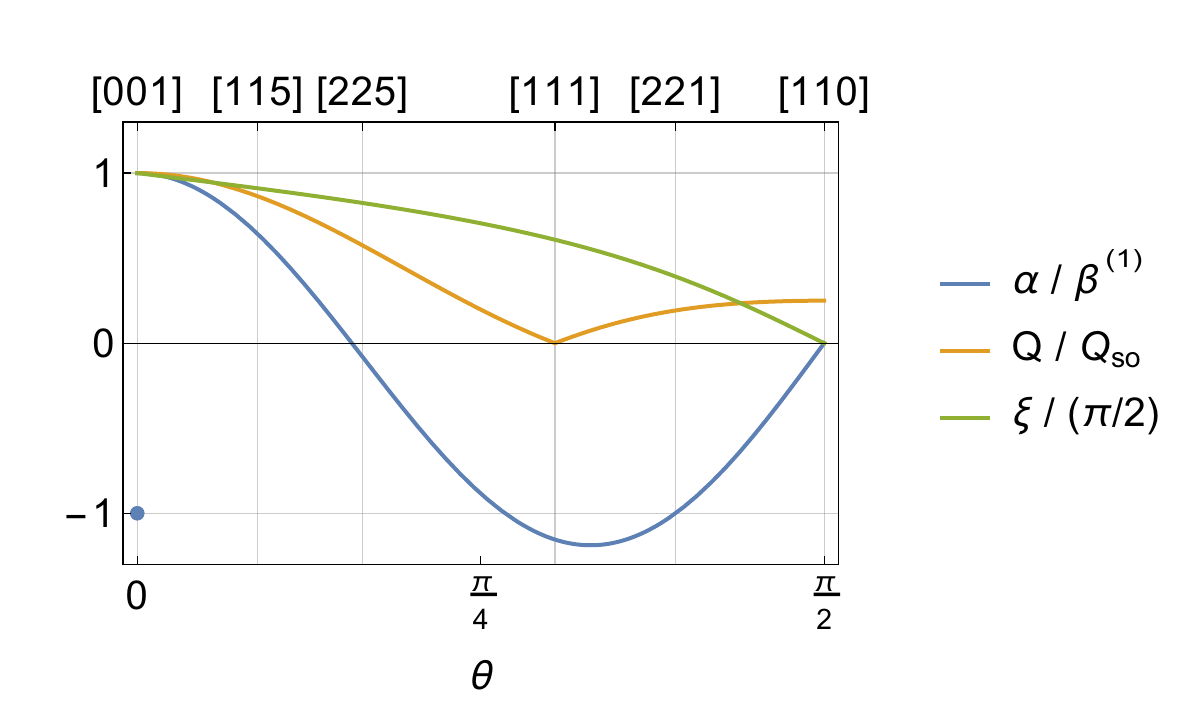}
  \caption{(Color online) Characteristic parameters in case of a persistent spin helix symmetry in dependence of the growth direction. Notice the degeneracy for $\alpha/\beta^{(1)}$ in the $[001]$ direction.}
  \label{plot:CSQ_1}
\end{figure}

 In Fig.~\ref{plot:CSQ_1}, we display the characteristic quantities in case of a persistent spin helix symmetry in dependence of the quantum well growth direction.
 Here, $\xi$ is defined as the polar angle between the surface normal $\mathbf{n}$ and the direction~$\boldsymbol\Pi$ of the collinear SOF.
 Obviously, $L_\text{s}$ reaches a minimum for a [001] orientation.
 For [111], i.e., $\eta=1/\sqrt{3}$, the wave vector vanishes due to an overall vanishing SOF as the Rashba and Dresselhaus contributions cancel each other exactly.
 Another peculiar situation occurs for $\eta=1/3$.
Similarly to a [110] 2DEG, it  yields a conserved spin quantity for a vanishing Rashba SOC.
As $\eta=1/3$ corresponds to an irrational Miller index, this growth direction  cannot be realized.
Yet, it can be well approximated by, e.g., a $[225]$ crystal vector.

\paragraph{Imprints on weak (anti)localization.} An indispensable tool to probe experimentally the  D'yakonov Perel' spin relaxation
are low-field magnetoconductivity (MC) measurements.
Quantum interference in weakly disordered conductors,  i.e., $\varepsilon_\text{F}\tau/\hbar\gg 1$, leads to a correction to the Drude conductivity $\Delta\sigma$ which is highly sensitive to magnetic fields as they break the  time-reversal invariance.
Depending on the strength and structure of the SOF the contribution to the conductivity can be positive or negative, which is denoted as weak localization (WL) or weak antilocalization (WAL), respectively.
For 2D electron systems the theory was developed by Hikami \textit{et al.}~\cite{nagaoka} and Iordanski \textit{et al.}~\cite{Iordanskii1994}.
However, Knap \textit{et al.}~\cite{Knap1996} discovered later that the spin relaxation rates induced by Rashba and Dresselhaus SOC are not additive and reflect the spin-preserving symmetries.
Subsequently, various theoretical models for WAL and WL have been successfully applied to planar and tubular 2DEGs~\cite{Schliemann2016,Kammermeier2016}.

Considering the standard white-noise model for the impurity potentials and weak disorder, we can write the 2D correction to the conductivity  as \cite{Kettemann2007a}
\begin{align}
\Delta\sigma={}&\frac{2e^2 }{h}\int\limits_{\mathcal{Q}<\sqrt{c_\text{e}}}\frac{d^2\mathcal{Q}}{(2\pi)^2}\bigg(\frac{1}{\mathcal{Q}^2+c_\phi+c_\text{B}}\notag\\
{}&-\sum_{j\in\{\pm1,0\}}\frac{1}{\lambda_j/Q_\text{so}^2+c_\phi+c_\text{B}}\bigg),
\label{conductivity}
\end{align}
with the conductance quantum  $2e^2/h$.
Moreover, we used the dimensionless orthogonal in-plane wave vectors of the 2DEG, $\boldsymbol{\mathcal{Q}}=(\mathcal{Q}_+,\mathcal{Q}_-)$ where $\mathcal{Q}=q/Q_\text{so}$.
Possible divergencies in the integral are removed by the upper and lower cutoffs $c_i={}1/(D_eQ_\text{so}^2\tau_i)$, $i \in \{\phi,\text{e},\text{B}\}$, due to finite dephasing, elastic scattering and magnetic phase shifting rates,  $\tau_\phi^{-1}$,  $\tau_\text{e}^{-1}$, and $\tau_\text{B}^{-1}$, respectively.
The latter takes into account external magnetic fields $\mathbf{B}=B\,\mathbf{n}$ perpendicular to the quantum well, i.e., $1/\tau_\text{B}=2 D_\text{e} e \vert B\vert / \hbar$~\cite{Beenakker1991}.
These fields are considered small enough that the Landau basis is not the appropriate choice.
The spectrum of the Cooperon and the spin diffusion equation are identical as far as time-reversal symmetry is not broken \cite{Malshukov1997}.
As a consequence, the spin relaxation rates, determined by the eigenvalues $\lambda_j$ of the spin diffusion operator $\Lambda_\text{sd}$, become manifested in the gaps of the triplet eigenvalues of the Cooperon and, thus, directly enter Eq.~(\ref{conductivity}).
In case of a gapless mode, that is, a vanishing spin relaxation, this results in a negative contribution to the conductivity, i.e., WL, despite of the presence of SOC and irrespective of its strength.
Therefore, a gate-controlled crossover from WAL to WL provides a solid evidence of spin-preserving symmetries~\cite{WalserNat2012,Kohda2012}.
The explicit form of $\Lambda_\text{sd}$, in case of two identical Miller indices, is given in the Supplemental Material.

\paragraph{Magnetoconductivity near SU(2) symmetry.} 
In the vicinity $\epsilon$ of the optimal ratio of Rashba and Dresselhaus SOC, i.e., $\alpha/\beta^{(1)}\mapsto\Gamma_0+\epsilon$, where $\epsilon \ll 1$, the structure of the eigenvalues $\lambda_j$, $j \in \{0,\pm1\}$, which are functions of the wave vector $\boldsymbol{\mathcal{Q}}$, can be approximated by three parabolas of the form
\begin{align}
\lambda_{j}/Q_\text{so}^2={}&\mathcal{Q}_+^2+\left(\mathcal{Q}_-+j
\zeta\right)^2+\Delta_{\vert j\vert}.
\end{align}
The minima of $\lambda_{\pm 1}$ are shifted to finite in-plane wave vectors $\mathcal{Q}_-=\pm\zeta$ which are oriented along $[\overline{1}10]$, representing the long-lived helical spin states.
Expanding $\Lambda_\text{sd}$ to lowest order in $\epsilon$ and neglecting all $\mathbf{q}$-independent terms yields a shift $\zeta^2\approx{}Q_0^2/Q_\text{so}^2+\Delta_0+\epsilon(1-3\eta^2)\sqrt{1-2\eta^2}$.
 Applying this and keeping only the leading terms  in $\epsilon$, one finds  $\Delta_0\approx{}2\Delta_1\approx\epsilon^2/4$.
The gaps  $\Delta_{\vert j\vert}=1/(D_eQ_\text{so}^2(\tau_\text{s})_{\vert j\vert})$ are a consequence of  the finite spin relaxation rates $(\tau_\text{s}^{-1})_{\vert j\vert}$ due to the broken SU(2) symmetry.
We stress that the gap at $\boldsymbol{\mathcal{Q}}=0$ is twice as large as the gap at $\boldsymbol{\mathcal{Q}}=(0,\pm\zeta)$. 
This fact is underlined by the results which are illustrated in Fig.~\ref{plot:minima}(b).
There, we compare the global minimum $\lambda_\text{min}$ of the spectrum of the spin diffusion operator $\Lambda_\text{sd}(\mathbf{q})$ with the one arising from the terms at $\mathbf{q}=0$ purely, Eq.~(\ref{eq:SRtensor}),
 for various growth directions.
Besides the cases of SU(2) symmetry (white lines), the minima at $\mathbf{q}=0$ are generally about a factor 2 larger than the minima $\lambda_\text{min}$.
These observations highlight the superior spin lifetime of helical spin densities which was previously observed in planar and tubular 2DEGs with Rashba SOC~\cite{Kettemann2007a,Kammermeier2016}. 
\begin{figure}
  \includegraphics[width= .9\columnwidth]{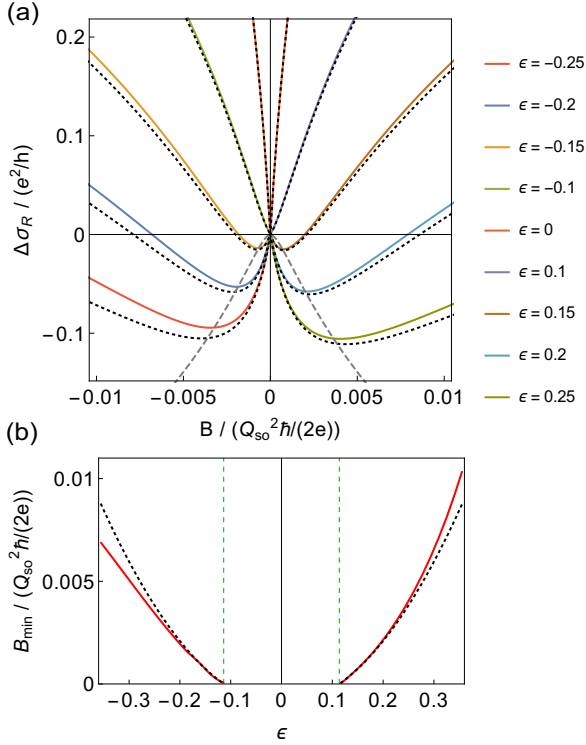}
  \caption{(Color online) 2DEG grown along [113] for $c_\text{e}/c_\phi=10^{3}$ and $c_\text{e}=1$ with $\alpha/\beta^{(1)}$ close to the SU(2) symmetry point. (a) Relative MC $\Delta\sigma_\text{R}(B)=\Delta\sigma(B)-\Delta\sigma(0)$ for different values of $\epsilon$. For compactness, we restrict the plots for $\epsilon<0$ ${(\epsilon>0)}$ to negative(positive) magnetic fields. The colored lines correspond to exact numerical calculations, the black dotted lines to the approximate expression, Eq.~(\ref{conductivity2}). Gray dashed lines show the trend of the minima $\Delta\sigma(B_\text{min})$ in dependence of $\epsilon$. (b) The respective MC  minimum as a function of  $\epsilon$. Red solid lines correspond to exact numerical calculations, black dotted and green dashed lines to approximate formulas.}
  \label{plot:wal}
\end{figure}

With this, the integral in Eq.~(\ref{conductivity}) yields an analytical result, which solely depends  on the quantities $\zeta$ and $\Delta_{0,1}$ and the cutoff parameters $c_i$, $i \in \{\phi,\text{e},\text{B}\}$:
\begin{align}
\Delta\sigma\approx{}&\frac{e^2 }{2\pi h}\ln \left(
\frac{4 \Upsilon_{100} \Upsilon_{010} \Upsilon_{001}^2 }    {\Upsilon_{000} \Upsilon_{110} \left( \Upsilon_{101} - \zeta^2 + \sqrt{\upsilon} \right)^2}
\right),
\label{conductivity2}
\end{align}
with the tensor $
\Upsilon_{jkl}={}c_\phi+c_\text{B}+j c_\text{e}+ k \Delta_0+l \Delta_1$ and $\upsilon={}\Upsilon_{101}^2+2\Upsilon_{\text{-}101}\zeta^2+\zeta^4$.
A particularly important  characteristic feature for experimental probing is the gate control of the MC minima $B_\text{min}$ where $\partial_{B}[\Delta\sigma(B)]=0$~\cite{Faniel2011,Yoshizumi2016}.
Exploiting the fact that $\epsilon$, $c_{\phi}$, $c_{\text{B}}$, and $c_\text{e}^{-1}$ are small quantities and neglecting the shift $\zeta$, we can use Eq.~(\ref{conductivity2}) to derive an approximate expression for $B_\text{min}$ as
\begin{align}
B_\text{min}\approx{}&\frac{(\sqrt{5}-1)m^2\widetilde{\alpha}^2}{2e\hbar^3}-\frac{\hbar}{2 eD_\text{e} \tau_\phi},
\label{Bmin}
\end{align}
where we defined $\widetilde{\alpha}=\beta^{(1)}\epsilon$.
According to this, the crossover from positive to negative MC  appears at $\widetilde{\alpha}^2\approx(1+\sqrt{5})\hbar^4/(4 D_\text{e} m^2\tau_\phi)$.
These simple relations allow for a direct determination of SOC coefficients and dephasing rate without parameter fitting. 
The quadratic scaling $B_\text{min}\propto\widetilde{\alpha}^2$ was recently confirmed in experiments~\cite{Faniel2011,Yoshizumi2016}.
We stress that our numerical investigations indicate that the approximate formulas show generally better agreement for $\epsilon>0$. 

To give an example, we consider a [113] orientated 2DEG, which has gathered attention as it facilitates long spin relaxation times of 2D hole systems~\cite{Ganichev2002}.
In Fig.~\ref{plot:wal}(a) we demonstrate the gate-induced crossover from positive to negative relative MC $\Delta\sigma_\text{R}(B)=\Delta\sigma(B)-\Delta\sigma(0)$ by varying the Rashba SOC strength, which is encapsulated in the quantity $\epsilon$, around the SU(2) symmetry point.
The colored lines correspond to the exact calculation by using Eqs.~(\ref{eq:spindiff}) and (\ref{conductivity}),  the black dotted lines to the approximate expression, Eq.~(\ref{conductivity2}).
The gray dashed lines depict the trend of the minimum $\Delta\sigma(B_\text{min})$  obtained by Eqs.~(\ref{conductivity2}) and (\ref{Bmin}).
The approximate formulas for the respective  $B_\text{min}$ and the horizontal offset (black dotted and green dashed lines) are compared to exact numerical calculation (red solid lines) in Fig.~\ref{plot:wal}(b).

In summary, we have identified general sufficient and necessary conditions for spin-preserving symmetries in 2DEGs of arbitrary growth directions. 
They demand a specific ratio of Rashba and Dresselhaus SOC for an arbitrary growth direction with at least two Miller indices equal in modulus.
Going from [001] to [110], the corresponding collinear SOF gradually transforms from in-plane to out-of-plane, simultaneously modifying the spin precession length.
Also, we determined two specific situations, i.e., [111] and [110], where the inclusion of higher angular harmonics of the Dresselhaus term continues to allow for a homogeneous persistent spin state.
Furthermore, by analyzing the spectrum of the spin diffusion equation, we show that besides the cases of perfect SU(2) symmetry, the spin of the long-lived homogeneous spin state relaxes about a factor two faster than for the helical spin state.
In addition, we derived analytical expressions for the magnetoconductivity and the location of its minimum around the SU(2) symmetry point.
The latter enables a fitting-free experimental determination of the transport parameters.
This work may trigger the interest for investigating 2DEGs with low-symmetry growth.
It opens up new perspectives and supports the progress towards tailoring  spintronic devices.

We thank Ch. Gradl, M. Schwemmer, and T. Korn for useful discussions. 
This work was supported by Deutsche Forschungsgemeinschaft via Grant No. SFB 689.



\section{Supplemental Material}\label{SM}
\subsection{Key requirement on crystal orientation for persistent spin states}
In the following, we prove that the realization of a SU(2) symmetry in a two-dimensional electron gas (2DEG) demands a growth direction with two Miller indices equal in modulus.

The Hamiltonian $\mathcal{H}_\text{SO}$ describing the spin-orbit coupling (SOC) for a 2DEG which is grown along an arbitrary normal unit vector $\mathbf{n}=(n_x,n_y,n_z)$  takes the form $\mathcal{H}_\text{SO}={}\boldsymbol{\Omega}(\mathbf{k})\cdot\boldsymbol{\sigma}$.
Focusing on the first angular harmonics, the spin-orbit field (SOF)  $\boldsymbol{\Omega}$ consists of  $\boldsymbol{\Omega}={}\boldsymbol{\Omega}_\text{R}+\boldsymbol{\Omega}_\text{D}^{(1)}$. The respective Rashba (R) and Dresselhaus (D) SOF, $\boldsymbol{\Omega}_\text{R}$ and $\boldsymbol{\Omega}_\text{D}^{(1)}$,
are defined  in Eq.~(2) of the main text.

The vector $\boldsymbol{\sigma}$ denotes the vector of Pauli matrices.
We can reformulate $\mathcal{H}_\text{SO}$ as
\begin{align}
\mathcal{H}_\text{SO}={}&\mathbf{k}^\top \Xi\; \boldsymbol{\sigma},
\end{align}
with a $k$-independent tensor $\Xi$ which collects the wave vector coefficients of the according components of the SOF.
In consequence of the 2D confinement, the wave vector obeys the relation $\mathbf{k}\cdot\mathbf{n}=0$.
Thus, without loss of generality we assume $n_z\neq 0$ and replace $k_z=-(k_x n_x+k_y n_y)/n_z$ in  $\boldsymbol{\Omega}$.
Using this and setting $\Gamma=\alpha/\beta^{(1)}$ gives
\begin{widetext}
\begin{align}
\Xi={}&\beta^{(1)}
\begin{pmatrix}
2 n_x^2+n_y^2-n_z^2+\Gamma  \frac{n_x n_y}{n_z} & -\left(4 n_x n_y +\Gamma \frac{n_x^2+n_z^2}{n_z}\right) & (n_y^2-n_x^2+2n_z^2)\frac{n_x}{n_z}+\Gamma n_y \\
4n_x n_y +\Gamma \frac{n_y^2+n_z^2}{n_z} & -\left(2n_y^2+n_x^2-n_z^2+\Gamma \frac{n_x n_y}{n_z}\right) & -\left((n_x^2-n_y^2+2n_z^2)\frac{n_y}{n_z}+\Gamma n_x\right) \\
0 & 0 & 0
\end{pmatrix}.
\end{align}
In case of a SU(2) symmetry, the SOC Hamiltonian $\mathcal{H}_\text{SO}$ can  be rewritten in the form $\mathcal{H}_\text{SO}=(\mathbf{k}\cdot\mathbf{Q})(\boldsymbol{\Pi}\cdot\boldsymbol{\sigma})$. 
In this formulation, both vectors $\mathbf{Q}$ and $\boldsymbol{\Pi}$ are required to be independent of  $k$.
The vector $\boldsymbol{\Pi}$ determines the direction of the collinear SOF and therefore the homogeneous persistent spin state.
In contrast, the vector $\mathbf{Q}$ induces the shift of the Fermi contours and describes the spin precession of the persistent spin helix.
Assuming this relation to hold, we can identify $\mathbf{a}_i=\mathbf{Q}\,\Pi_i$,  $i \in \{1,2,3\}$, where $\mathbf{a}_i$ denotes the $i$-th column vector of $\Xi$.
Hence, in order to obtain SU(2) symmetry, the column vectors $\mathbf{a}_i$ of $\Xi$ need to be collinear.
This yields three equations $\mathbf{a}_1\times\mathbf{a}_2=\mathbf{a}_3\times\mathbf{a}_1=\mathbf{a}_2\times\mathbf{a}_3=0$ which are equivalent to 
\begin{align}
n_x\Gamma^2+n_yn_z(10n_x^2+n_y^2+n_z^2)\Gamma={}&n_x[n_x^4+2(n_y^4+n_z^4)-3n_x^2(n_y^2+n_z^2)-11n_y^2n_z^2],\label{eq1}\\
n_y\Gamma^2+n_zn_x(10n_y^2+n_z^2+n_x^2)\Gamma={}&n_y[n_y^4+2(n_z^4+n_x^4)-3n_y^2(n_z^2+n_x^2)-11n_z^2n_x^2],\label{eq2}\\
n_z\Gamma^2+n_xn_y(10n_z^2+n_x^2+n_y^2)\Gamma={}&n_z[n_z^4+2(n_x^4+n_y^4)-3n_z^2(n_x^2+n_y^2)-11n_x^2n_y^2].\label{eq3}
\end{align}
\end{widetext}
From $n_x=0$ in Eq.~(\ref{eq1}) follows that either $\Gamma=0$ or $n_y=0$.
The latter case corresponds to two equal indices.
Using the solution $n_x=\Gamma=0$ in Eq.~(\ref{eq2}) leads again to $n_y=0$.
As this allows to exclude the cases $n_x=n_y=0$ from the discussion, we can eliminate $\Gamma^2$ by subtracting the distinct equations from each other.
Taking additionally into account normalization, i.e., $n_x^2+n_y^2+n_z^2=1$,  one finds
\begin{align}
\Gamma(n_x^2-n_y^2)={}&\frac{n_x n_y}{n_z}(1-9n_z^2)(n_x^2-n_y^2),\\
\Gamma(n_y^2-n_z^2)={}&\frac{n_y n_z}{n_x}(1-9n_x^2)(n_y^2-n_z^2),\\
\Gamma(n_z^2-n_x^2)={}&\frac{n_z n_x}{n_y}(1-9n_y^2)(n_z^2-n_x^2).
\end{align}
On condition that all indices $\vert n_i \vert$ are different from each other, we can cancel the factors $(n_i^2-n_j^2)$.
This, however, causes inconsistent solutions for $\Gamma$.
As a result, at least two indices are required to be equal in modulus and we obtain three cases:
\begin{align}
n_x=\pm n_y\: :\:\Gamma={}&\pm n_z - 9 n_x n_y n_z,\\
n_y=\pm n_z\: :\:\Gamma={}&\pm n_x - 9 n_x n_y n_z,\\
n_z=\pm n_x\: :\:\Gamma={}&\pm n_y - 9 n_x n_y n_z,
\end{align}
which includes the result of the main text.

In the following section, we demonstrate that,  in turn, starting from the assumption $n_x=n_y=\eta$, the relation $\Gamma=(1-9\eta^2)n_z$ must hold true in order to generate a collinear spin-orbit field.
The latter induces SU(2) symmetry since it allows to reformulate the SOC Hamiltonian $\mathcal{H}_\text{SO}$ in the form $\mathcal{H}_\text{SO}=(\mathbf{k}\cdot\mathbf{Q})(\boldsymbol{\Pi}\cdot\boldsymbol{\sigma})$ 
as it is shown in the main text.

\subsection{Collinear spin-orbit field and the effects of higher angular harmonics}
\label{sec:A1}
Again, we assume a 2DEG which is oriented along an arbitrary normal unit vector $\mathbf{n}=(n_x,n_y,n_z)$.
The Hamiltonian $\mathcal{H}_\text{SO}$ which describes the SOC  is written as 
\begin{align}
\mathcal{H}_\text{SO}={}&\boldsymbol{\Omega}\cdot\boldsymbol{\sigma}=(\boldsymbol{\Omega}_\text{R}+\boldsymbol{\Omega}_\text{D})\cdot\boldsymbol{\sigma},
\label{eq:dre}
\end{align}
where the Rashba SOF $\boldsymbol{\Omega}_\text{R}$ is defined in Eq.~(2) of the main text.
The full Dresselhaus SOF $\boldsymbol{\Omega}_\text{D}$, including angular harmonics up to third order, is given by $\boldsymbol{\Omega}_\text{D}=\boldsymbol{\Omega}_\text{D}^{(1)}+\boldsymbol{\Omega}_\text{D}^{(3)}={}\gamma_\text{D}\boldsymbol{\nu}$ with the components \cite{Dyakonov1986,Zutic2004a}
\begin{align}
\nu_x={}&\left(\frac{\pi}{a}\right)^2\left[2n_x(n_yk_y-n_zk_z)+k_x(n_y^2-n_z^2)\right]\notag\\
&+k_x\left(k_y^2-k_z^2\right)
\label{eq:dre_comp}
\end{align}
and similar for $\nu_y$ and $\nu_z$ by cyclic index permutation. 
The vector $\boldsymbol{\sigma}$ denotes the vector of Pauli matrices, $a$ the width of the square well confinement along the growth direction, and $\gamma_\text{D}$ a material and confinement specific parameter.
In consequence of the 2D confinement, the wave vectors obey the relation $\mathbf{k}\cdot\mathbf{n}=0$.
In the given representation, the basis vectors $\boldsymbol{\hat{x}}$, $\boldsymbol{\hat{y}}$, and $\boldsymbol{\hat{z}}$ correspond to the principal crystal axes [100], [010], and [001], respectively.

It is convenient to rotate the Hamiltonian $\mathcal{H}_\text{SO}$ such that the $z$-axis of the transformed system is aligned with the 2DEG's growth direction. 
Focusing on the scenario where the persistent spin helix symmetry can be realized up to higher angular harmonics, we set $n_x=n_y\equiv \eta$ and $n_z=\sqrt{1-2\eta^2}$.
For simplicity, we restrict to the first octant only, i.e., $\eta \in \left[0,1/\sqrt{2}\right]$. 
The relation to the polar angle $\theta$ with respect to $[001]$ yields $\eta=\sin(\theta)/\sqrt{2}$.
With this, the rotation can be performed by means of the rotation matrix 
\begin{align}
\mathcal{R}={}&\frac{1}{\sqrt{2}}
\begin{pmatrix}
n_z&-1 &\sqrt{2}\eta \\
n_z&1 &\sqrt{2}\eta \\
-2\eta &0 &\sqrt{2}n_z \\
\end{pmatrix}.
\label{eq:rotmatrix}
\end{align}
The Hamiltonian in the rotated system, i.e., $\mathcal{H}_\text{SO}'={\boldsymbol{\Omega}'\cdot\boldsymbol{\sigma}'}$, is obtained by replacing $\mathbf{k}\mapsto\mathcal{R}\cdot\mathbf{k}'$ and $\boldsymbol{\sigma}\mapsto\mathcal{R}\cdot\boldsymbol{\sigma}'$ with the according basis vectors $\boldsymbol{\hat{x}}'=(n_z,n_z,-2\eta)/\sqrt{2}$, $\boldsymbol{\hat{y}}'=(-1,1,0)/\sqrt{2}$, and $\boldsymbol{\hat{z}}'=(\eta,\eta,n_z)$.

After applying this transformation, the Rashba SOF lies in the plane of the quantum well and reads
\begin{align}
\boldsymbol{\Omega}_\text{R}'={}&\alpha k' \;(\sin(\varphi),-\cos(\varphi),0),
\end{align}
with $\alpha=\gamma_\text{R}\mathcal{E}_0$ where $\gamma_\text{R}$  is a material and confinement specific parameter and $\mathcal{E}_0$ results from an electric field $\boldsymbol{\mathcal{E}}=\mathcal{E}_0\boldsymbol{\hat{z}}'$ perpendicular to the 2DEG.
Here, we introduced polar coordinates for the in-plane wave vectors, $k_x'=k' \cos (\varphi)$ and $k_y'=k'\sin (\varphi)$.
The Dresselhaus SOF  $\boldsymbol{\Omega}_\text{D}'$ can be split into two contributions that contain the angular harmonics in $\mathbf{k}$ of the first and third order,  $\boldsymbol{\Omega}_\text{D}'^{(1)}$ and $\boldsymbol{\Omega}_\text{D}'^{(3)}$, respectively.
Thus, we find
\begin{align}
\boldsymbol{\Omega}_\text{D}'={}&\boldsymbol{\Omega}_\text{D}'^{(1)}+\boldsymbol{\Omega}_\text{D}'^{(3)},
\label{eq:dre_sof}
\end{align}
with the SOF w.r.t. the $l$-th angular harmonics
\begin{align}
\boldsymbol{\Omega}_\text{D}'^{(l)}={}&\beta^{(l)}k'
\begin{pmatrix}
b_1^{(l)}\sin(l\varphi)\\
b_2^{(l)}\cos(l\varphi)\\
b_3^{(l)}\sin(l\varphi))
\end{pmatrix},
\end{align}
where  $\beta^{(1)}=\gamma_\text{D}\left[(\pi/a)^2-k'^2/4\right]$, $\beta^{(3)}=\gamma_\text{D}k'^2/4$, and the respective coefficients  $b_j^{(l)}$ are comprised in the vectors
\begin{align}
\mathbf{b}^{(1)}={}&
\begin{pmatrix}
(1+3\eta^2)n_z\\
(1-9\eta^2)n_z\\
-\sqrt{2}\eta(1-3\eta^2)
\end{pmatrix},\:
\mathbf{b}^{(3)}=
\begin{pmatrix}
(1-3\eta^2)n_z\\
-(1-3\eta^2)n_z\\
3\sqrt{2}\eta(1-\eta^2)
\end{pmatrix}.
\end{align}
We note that each of the Dresselhaus fields lies in a plane which is defined by the corresponding normal vector $\mathbf{v}'^{(l)}=(b_3^{(l)},0,-b_1^{(l)})$.
The planes coincide if $\eta=0$ or $\eta=1/\sqrt{2}$, i.e., in case of a [001] or [110] 2DEG. 
In general, the field  $\boldsymbol{\Omega}_\text{D}'^{(3)}$ gives a correction $\boldsymbol{\Omega}_\text{D}^{(3)}\cdot\boldsymbol{\sigma}$ to the Hamiltonian $\mathcal{H}$ in Eqs.~(1) and (6) of the main text.
To this end, the Dresselhaus contribution due to third angular harmonics $\boldsymbol{\Omega}_\text{D}'^{(3)}\cdot\boldsymbol{\sigma}'$ has to be back transformed to the initial coordinate system corresponding to the principal crystal axes.
This is achieved by replacing  $\boldsymbol{\sigma}'\mapsto\mathcal{R}^{-1}\boldsymbol{\sigma}$, $\mathbf{k}'\mapsto\mathcal{R}^{-1}\mathbf{k}$, and using the relations $k'\cos(3\varphi)=k_x'(k_x'^2-3k_y'^2)$ and $k'\sin(3\varphi)=-k_y'(k_y'^2-3k_x'^2)$.

Typically, taking into account the effects of the third angular harmonics  inhibits the realization of a perfect SU(2) symmetry.
The spin rotation is no more \textit{well-defined} as it depends on the electron's propagated path.
Nonetheless, in the following we show that the [111] and [110] directions allow to construct a collinear SOF despite of the presence of $\boldsymbol{\Omega_\text{D}}^{(3)}$ which facilitates homogeneous persistent spin states.
By adding the contributions $\boldsymbol{\Omega}_\text{R}'+\boldsymbol{\Omega}_\text{D}'^{(1)}$ it becomes obvious that a collinear SOF field is formed if the Rashba and Dresselhaus coefficients $\alpha$ and $\beta^{(1)}$ fulfill the relation
\begin{align}
\alpha/\beta^{(1)}=b_2^{(1)}=(1-9\eta^2)\sqrt{1-2\eta^2}
\end{align}
which generates the SU(2) symmetry as demonstrated in the main text. 
Adopting this and focusing on the first angular harmonics yields the collinear field 
\begin{align}
\boldsymbol{\Omega}_\text{R}'+\boldsymbol{\Omega}_\text{D}'^{(1)}={}&\beta^{(1)} k' \sin(\varphi)(3\eta^2-1)
\begin{pmatrix}
-2n_z\\
0\\
\sqrt{2}\eta
\end{pmatrix}.
\end{align}
Comparing with $\boldsymbol{\Omega}_\text{D}'^{(3)}$, we observe that the collinearity is generally destroyed apart from two particular situations where $\eta=1/\sqrt{3}$ or $\eta=1/\sqrt{2}$, that is, the [111] or [110] direction, respectively.
For [111], the SOF spanned by $\boldsymbol{\Omega}_\text{R}'+\boldsymbol{\Omega}_\text{D}'^{(1)}$ vanishes as both contributions cancel each other. The remaining field is given by $\boldsymbol{\Omega}_\text{D}'^{(3)}$ which is collinear to the [111] direction.
In case of a [110] 2DEG, we have $\boldsymbol{\Omega}_\text{R}'=0$ and the $\boldsymbol{\Omega}_\text{D}'^{(3)}$ contribution is collinear to the [110] axis which coincides with the collinear field due to $\boldsymbol{\Omega}_\text{D}'^{(1)}$.

Summarizing, we have demonstrated that choosing a growth direction which corresponds to identical Miller indices and focusing on the first angular harmonics opens the possibility to generate a collinear SOF.
Also, we identified two specific scenarios, where the inclusion of higher angular harmonics does not destroy this collinearity and continues to allows for a homogeneous persistent spin state.

\subsection{Spin diffusion operator for two identical Miller indices}
Let us consider again a 2DEG grown along a crystal direction with two identical Miller indices with focus on the first octant, i.e., $n_x=n_y\equiv \eta$ and $n_z=\sqrt{1-2\eta^2}$ for $\eta \in \left[0,1/\sqrt{2}\right]$.
Analogously to the previous section, by replacing $\mathbf{q}\mapsto\mathcal{R}\cdot\mathbf{q}'$ we choose the in-plane coordinate representation.
Furthermore, we define the dimensionless in-plane wave vectors of the spin density as $\mathcal{Q}_+=\mathcal{Q} \cos(\varphi)$ and $\mathcal{Q}_-=\mathcal{Q} \sin(\varphi)$ with $\mathcal{Q}=q'/Q_\text{so}$ which correspond to the basis vectors $\boldsymbol{\hat{x}}'=(n_z,n_z,-2\eta)/\sqrt{2}$ and $\boldsymbol{\hat{y}}'=(-1,1,0)/\sqrt{2}$, respectively.
With these definitions, the spin diffusion operator can be written as
\begin{align}
\Lambda_\text{sd}/Q_\text{so}^2={}&
\begin{pmatrix}
K&L&M\\
L^*&N&O\\
M^*&O^*&P
\end{pmatrix},
\end{align}
with the components
\begin{align}
K={}&\mathcal{Q}^2+\frac{1}{4}\left[1+\Gamma^2+ \left(\Gamma^2+ 16 \Gamma n_z -3 \right)\eta^2\right.\notag\\
&\left.+43\eta^4-81\eta^6\right],\\
L={}&i\sqrt{2} \mathcal{Q} \sin (\varphi) \eta(2 n_z + \Gamma) -\frac{1}{2}n_z \Gamma \notag\\
& + \frac{1}{4}\left[ (\Gamma^2+ 10 n_z \Gamma-9 ) \eta^2 +58 \eta^4 - 81 \eta^6\right],\\
M={}&\frac{\eta }{4}[\Gamma (3 - 5 \eta^2)+n_z (2 + \Gamma^2 - 2 \eta^2)]\notag\\
&- i \frac{\mathcal{Q}}{\sqrt{2}}\left\{ [\Gamma + n_z (9 \eta^2-1 )] \cos(\varphi)\right.\notag\\
&\left. + (\eta^2-n_z \Gamma-1) \sin(\varphi)\right\},\\
N={}&K,\\
O={}&M-i\sqrt{2}\mathcal{Q}\sin(\varphi)(1+n_z\Gamma-\eta^2),\\
P={}&K+\frac{1}{4}(1-3\eta^2)(1 + \Gamma^2 - 16 \eta^2 + 27 \eta^4),
\end{align}
where $\Gamma=\alpha/\beta^{(1)}$, $Q_\text{so}=4m\beta^{(1)}/\hbar^2$, and higher angular harmonics $\propto\beta^{(3)}$ are neglected.
The eigenvalues $\lambda_j$ of $\Lambda_\text{sd}$ directly enter  Eq.~(9) of the main text which yields the weak (anti)localization correction to the Drude conductivity.

\bibliographystyle{apsrev4-1}
\bibliography{WK}

\def\url#1{}
\begin{thebibliography}{30}%
\makeatletter
\providecommand \@ifxundefined [1]{%
 \@ifx{#1\undefined}
}%
\providecommand \@ifnum [1]{%
 \ifnum #1\expandafter \@firstoftwo
 \else \expandafter \@secondoftwo
 \fi
}%
\providecommand \@ifx [1]{%
 \ifx #1\expandafter \@firstoftwo
 \else \expandafter \@secondoftwo
 \fi
}%
\providecommand \natexlab [1]{#1}%
\providecommand \enquote  [1]{``#1''}%
\providecommand \bibnamefont  [1]{#1}%
\providecommand \bibfnamefont [1]{#1}%
\providecommand \citenamefont [1]{#1}%
\providecommand \href@noop [0]{\@secondoftwo}%
\providecommand \href [0]{\begingroup \@sanitize@url \@href}%
\providecommand \@href[1]{\@@startlink{#1}\@@href}%
\providecommand \@@href[1]{\endgroup#1\@@endlink}%
\providecommand \@sanitize@url [0]{\catcode `\\12\catcode `\$12\catcode
  `\&12\catcode `\#12\catcode `\^12\catcode `\_12\catcode `\%12\relax}%
\providecommand \@@startlink[1]{}%
\providecommand \@@endlink[0]{}%
\providecommand \url  [0]{\begingroup\@sanitize@url \@url }%
\providecommand \@url [1]{\endgroup\@href {#1}{\urlprefix }}%
\providecommand \urlprefix  [0]{URL }%
\providecommand \Eprint [0]{\href }%
\providecommand \doibase [0]{http://dx.doi.org/}%
\providecommand \selectlanguage [0]{\@gobble}%
\providecommand \bibinfo  [0]{\@secondoftwo}%
\providecommand \bibfield  [0]{\@secondoftwo}%
\providecommand \translation [1]{[#1]}%
\providecommand \BibitemOpen [0]{}%
\providecommand \bibitemStop [0]{}%
\providecommand \bibitemNoStop [0]{.\EOS\space}%
\providecommand \EOS [0]{\spacefactor3000\relax}%
\providecommand \BibitemShut  [1]{\csname bibitem#1\endcsname}%
\let\auto@bib@innerbib\@empty
\bibitem [{\citenamefont {Awschalom}\ and\ \citenamefont
  {Flatte}(2007)}]{Awschalom2007}%
  \BibitemOpen
  \bibfield  {author} {\bibinfo {author} {\bibfnamefont {D.~D.}\ \bibnamefont
  {Awschalom}}\ and\ \bibinfo {author} {\bibfnamefont {M.~E.}\ \bibnamefont
  {Flatte}},\ }\href {http://dx.doi.org/10.1038/nphys551} {\bibfield  {journal}
  {\bibinfo  {journal} {Nat. Phys.}\ }\textbf {\bibinfo {volume} {3}},\
  \bibinfo {pages} {153} (\bibinfo {year} {2007})}\BibitemShut {NoStop}%
\bibitem [{\citenamefont {Schliemann}\ \emph {et~al.}(2003)\citenamefont
  {Schliemann}, \citenamefont {Egues},\ and\ \citenamefont
  {Loss}}]{Schliemann2003}%
  \BibitemOpen
  \bibfield  {author} {\bibinfo {author} {\bibfnamefont {J.}~\bibnamefont
  {Schliemann}}, \bibinfo {author} {\bibfnamefont {J.~C.}\ \bibnamefont
  {Egues}}, \ and\ \bibinfo {author} {\bibfnamefont {D.}~\bibnamefont {Loss}},\
  }\href {\doibase 10.1103/PhysRevLett.90.146801} {\bibfield  {journal}
  {\bibinfo  {journal} {Phys. Rev. Lett.}\ }\textbf {\bibinfo {volume} {90}},\
  \bibinfo {pages} {146801} (\bibinfo {year} {2003})}\BibitemShut {NoStop}%
\bibitem [{\citenamefont {Bernevig}\ \emph {et~al.}(2006)\citenamefont
  {Bernevig}, \citenamefont {Orenstein},\ and\ \citenamefont
  {Zhang}}]{Bernevig2006}%
  \BibitemOpen
  \bibfield  {author} {\bibinfo {author} {\bibfnamefont {B.~A.}\ \bibnamefont
  {Bernevig}}, \bibinfo {author} {\bibfnamefont {J.}~\bibnamefont {Orenstein}},
  \ and\ \bibinfo {author} {\bibfnamefont {S.~C.}\ \bibnamefont {Zhang}},\
  }\href {\doibase 10.1103/PhysRevLett.97.236601} {\bibfield  {journal}
  {\bibinfo  {journal} {Phys. Rev. Lett.}\ }\textbf {\bibinfo {volume} {97}},\
  \bibinfo {pages} {236601} (\bibinfo {year} {2006})}\BibitemShut {NoStop}%
\bibitem [{\citenamefont {Trushin}\ and\ \citenamefont
  {Schliemann}(2007)}]{Trushin2007}%
  \BibitemOpen
  \bibfield  {author} {\bibinfo {author} {\bibfnamefont {M.}~\bibnamefont
  {Trushin}}\ and\ \bibinfo {author} {\bibfnamefont {J.}~\bibnamefont
  {Schliemann}},\ }\href {http://stacks.iop.org/1367-2630/9/i=9/a=346}
  {\bibfield  {journal} {\bibinfo  {journal} {New J. Phys.}\ }\textbf {\bibinfo
  {volume} {9}},\ \bibinfo {pages} {346} (\bibinfo {year} {2007})}\BibitemShut
  {NoStop}%
\bibitem [{\citenamefont {Sacksteder}\ and\ \citenamefont
  {Bernevig}(2014)}]{Sacksteder2014}%
  \BibitemOpen
  \bibfield  {author} {\bibinfo {author} {\bibfnamefont {V.~E.}\ \bibnamefont
  {Sacksteder}}\ and\ \bibinfo {author} {\bibfnamefont {B.~A.}\ \bibnamefont
  {Bernevig}},\ }\href {\doibase 10.1103/PhysRevB.89.161307} {\bibfield
  {journal} {\bibinfo  {journal} {Phys. Rev. B}\ }\textbf {\bibinfo {volume}
  {89}},\ \bibinfo {pages} {161307} (\bibinfo {year} {2014})}\BibitemShut
  {NoStop}%
\bibitem [{\citenamefont {Dollinger}\ \emph {et~al.}(2014)\citenamefont
  {Dollinger}, \citenamefont {Kammermeier}, \citenamefont {Scholz},
  \citenamefont {Wenk}, \citenamefont {Schliemann}, \citenamefont {Richter},\
  and\ \citenamefont {Winkler}}]{Dollinger2014}%
  \BibitemOpen
  \bibfield  {author} {\bibinfo {author} {\bibfnamefont {T.}~\bibnamefont
  {Dollinger}}, \bibinfo {author} {\bibfnamefont {M.}~\bibnamefont
  {Kammermeier}}, \bibinfo {author} {\bibfnamefont {A.}~\bibnamefont {Scholz}},
  \bibinfo {author} {\bibfnamefont {P.}~\bibnamefont {Wenk}}, \bibinfo {author}
  {\bibfnamefont {J.}~\bibnamefont {Schliemann}}, \bibinfo {author}
  {\bibfnamefont {K.}~\bibnamefont {Richter}}, \ and\ \bibinfo {author}
  {\bibfnamefont {R.}~\bibnamefont {Winkler}},\ }\href {\doibase
  10.1103/PhysRevB.90.115306} {\bibfield  {journal} {\bibinfo  {journal} {Phys.
  Rev. B}\ }\textbf {\bibinfo {volume} {90}},\ \bibinfo {pages} {115306}
  (\bibinfo {year} {2014})}\BibitemShut {NoStop}%
\bibitem [{\citenamefont {Wenk}\ \emph {et~al.}(2016)\citenamefont {Wenk},
  \citenamefont {Kammermeier},\ and\ \citenamefont {Schliemann}}]{Wenk2016}%
  \BibitemOpen
  \bibfield  {author} {\bibinfo {author} {\bibfnamefont {P.}~\bibnamefont
  {Wenk}}, \bibinfo {author} {\bibfnamefont {M.}~\bibnamefont {Kammermeier}}, \
  and\ \bibinfo {author} {\bibfnamefont {J.}~\bibnamefont {Schliemann}},\
  }\href {\doibase 10.1103/PhysRevB.93.115312} {\bibfield  {journal} {\bibinfo
  {journal} {Phys. Rev. B}\ }\textbf {\bibinfo {volume} {93}},\ \bibinfo
  {pages} {115312} (\bibinfo {year} {2016})}\BibitemShut {NoStop}%
\bibitem [{\citenamefont {Koralek}\ \emph {et~al.}(2009)\citenamefont
  {Koralek}, \citenamefont {Weber}, \citenamefont {Orenstein}, \citenamefont
  {Bernevig}, \citenamefont {Zhang}, \citenamefont {Mack},\ and\ \citenamefont
  {Awschalom}}]{Koralek2009}%
  \BibitemOpen
  \bibfield  {author} {\bibinfo {author} {\bibfnamefont {J.~D.}\ \bibnamefont
  {Koralek}}, \bibinfo {author} {\bibfnamefont {C.~P.}\ \bibnamefont {Weber}},
  \bibinfo {author} {\bibfnamefont {J.}~\bibnamefont {Orenstein}}, \bibinfo
  {author} {\bibfnamefont {B.~A.}\ \bibnamefont {Bernevig}}, \bibinfo {author}
  {\bibfnamefont {S.~C.}\ \bibnamefont {Zhang}}, \bibinfo {author}
  {\bibfnamefont {S.}~\bibnamefont {Mack}}, \ and\ \bibinfo {author}
  {\bibfnamefont {D.~D.}\ \bibnamefont {Awschalom}},\ }\href
  {http://dx.doi.org/10.1038/nature07871} {\bibfield  {journal} {\bibinfo
  {journal} {Nature}\ }\textbf {\bibinfo {volume} {458}},\ \bibinfo {pages}
  {610} (\bibinfo {year} {2009})}\BibitemShut {NoStop}%
\bibitem [{\citenamefont {Kunihashi}\ \emph {et~al.}(2009)\citenamefont
  {Kunihashi}, \citenamefont {Kohda},\ and\ \citenamefont
  {Nitta}}]{Kunihashi2009}%
  \BibitemOpen
  \bibfield  {author} {\bibinfo {author} {\bibfnamefont {Y.}~\bibnamefont
  {Kunihashi}}, \bibinfo {author} {\bibfnamefont {M.}~\bibnamefont {Kohda}}, \
  and\ \bibinfo {author} {\bibfnamefont {J.}~\bibnamefont {Nitta}},\ }\href
  {\doibase 10.1103/PhysRevLett.102.226601} {\bibfield  {journal} {\bibinfo
  {journal} {Phys. Rev. Lett.}\ }\textbf {\bibinfo {volume} {102}},\ \bibinfo
  {pages} {226601 } (\bibinfo {year} {2009})}\BibitemShut {NoStop}%
\bibitem [{\citenamefont {Faniel}\ \emph {et~al.}(2011)\citenamefont {Faniel},
  \citenamefont {Matsuura}, \citenamefont {Mineshige}, \citenamefont {Sekine},\
  and\ \citenamefont {Koga}}]{Faniel2011}%
  \BibitemOpen
  \bibfield  {author} {\bibinfo {author} {\bibfnamefont {S.}~\bibnamefont
  {Faniel}}, \bibinfo {author} {\bibfnamefont {T.}~\bibnamefont {Matsuura}},
  \bibinfo {author} {\bibfnamefont {S.}~\bibnamefont {Mineshige}}, \bibinfo
  {author} {\bibfnamefont {Y.}~\bibnamefont {Sekine}}, \ and\ \bibinfo {author}
  {\bibfnamefont {T.}~\bibnamefont {Koga}},\ }\href {\doibase
  10.1103/PhysRevB.83.115309} {\bibfield  {journal} {\bibinfo  {journal} {Phys.
  Rev. B}\ }\textbf {\bibinfo {volume} {83}},\ \bibinfo {pages} {115309}
  (\bibinfo {year} {2011})}\BibitemShut {NoStop}%
\bibitem [{\citenamefont {Walser}\ \emph {et~al.}(2012)\citenamefont {Walser},
  \citenamefont {Reichl}, \citenamefont {Wegscheider},\ and\ \citenamefont
  {Salis}}]{WalserNat2012}%
  \BibitemOpen
  \bibfield  {author} {\bibinfo {author} {\bibfnamefont {M.~P.}\ \bibnamefont
  {Walser}}, \bibinfo {author} {\bibfnamefont {C.}~\bibnamefont {Reichl}},
  \bibinfo {author} {\bibfnamefont {W.}~\bibnamefont {Wegscheider}}, \ and\
  \bibinfo {author} {\bibfnamefont {G.}~\bibnamefont {Salis}},\ }\href
  {\doibase http://dx.doi.org/10.1038/nphys2383} {\bibfield  {journal}
  {\bibinfo  {journal} {Nat. Phys.}\ }\textbf {\bibinfo {volume} {8}},\
  \bibinfo {pages} {757} (\bibinfo {year} {2012})}\BibitemShut {NoStop}%
\bibitem [{\citenamefont {Kohda}\ \emph {et~al.}(2012)\citenamefont {Kohda},
  \citenamefont {Lechner}, \citenamefont {Kunihashi}, \citenamefont
  {Dollinger}, \citenamefont {Olbrich}, \citenamefont {Sch\"onhuber},
  \citenamefont {Caspers}, \citenamefont {Bel'kov}, \citenamefont {Golub},
  \citenamefont {Weiss}, \citenamefont {Richter}, \citenamefont {Nitta},\ and\
  \citenamefont {Ganichev}}]{Kohda2012}%
  \BibitemOpen
  \bibfield  {author} {\bibinfo {author} {\bibfnamefont {M.}~\bibnamefont
  {Kohda}}, \bibinfo {author} {\bibfnamefont {V.}~\bibnamefont {Lechner}},
  \bibinfo {author} {\bibfnamefont {Y.}~\bibnamefont {Kunihashi}}, \bibinfo
  {author} {\bibfnamefont {T.}~\bibnamefont {Dollinger}}, \bibinfo {author}
  {\bibfnamefont {P.}~\bibnamefont {Olbrich}}, \bibinfo {author} {\bibfnamefont
  {C.}~\bibnamefont {Sch\"onhuber}}, \bibinfo {author} {\bibfnamefont
  {I.}~\bibnamefont {Caspers}}, \bibinfo {author} {\bibfnamefont {V.~V.}\
  \bibnamefont {Bel'kov}}, \bibinfo {author} {\bibfnamefont {L.~E.}\
  \bibnamefont {Golub}}, \bibinfo {author} {\bibfnamefont {D.}~\bibnamefont
  {Weiss}}, \bibinfo {author} {\bibfnamefont {K.}~\bibnamefont {Richter}},
  \bibinfo {author} {\bibfnamefont {J.}~\bibnamefont {Nitta}}, \ and\ \bibinfo
  {author} {\bibfnamefont {S.~D.}\ \bibnamefont {Ganichev}},\ }\href {\doibase
  10.1103/PhysRevB.86.081306} {\bibfield  {journal} {\bibinfo  {journal} {Phys.
  Rev. B}\ }\textbf {\bibinfo {volume} {86}},\ \bibinfo {pages} {081306}
  (\bibinfo {year} {2012})}\BibitemShut {NoStop}%
\bibitem [{\citenamefont {Ishihara}\ \emph {et~al.}(2014)\citenamefont
  {Ishihara}, \citenamefont {Ohno},\ and\ \citenamefont {Ohno}}]{Ishihara2014}%
  \BibitemOpen
  \bibfield  {author} {\bibinfo {author} {\bibfnamefont {J.}~\bibnamefont
  {Ishihara}}, \bibinfo {author} {\bibfnamefont {Y.}~\bibnamefont {Ohno}}, \
  and\ \bibinfo {author} {\bibfnamefont {H.}~\bibnamefont {Ohno}},\ }\href
  {http://stacks.iop.org/1882-0786/7/i=1/a=013001} {\bibfield  {journal}
  {\bibinfo  {journal} {Appl. Phys. Express}\ }\textbf {\bibinfo {volume}
  {7}},\ \bibinfo {pages} {013001} (\bibinfo {year} {2014})}\BibitemShut
  {NoStop}%
\bibitem [{\citenamefont {Yoshizumi}\ \emph {et~al.}(2016)\citenamefont
  {Yoshizumi}, \citenamefont {Sasaki}, \citenamefont {Kohda},\ and\
  \citenamefont {Nitta}}]{Yoshizumi2016}%
  \BibitemOpen
  \bibfield  {author} {\bibinfo {author} {\bibfnamefont {K.}~\bibnamefont
  {Yoshizumi}}, \bibinfo {author} {\bibfnamefont {A.}~\bibnamefont {Sasaki}},
  \bibinfo {author} {\bibfnamefont {M.}~\bibnamefont {Kohda}}, \ and\ \bibinfo
  {author} {\bibfnamefont {J.}~\bibnamefont {Nitta}},\ }\href
  {http://scitation.aip.org/content/aip/journal/apl/108/13/10.1063/1.4944931}
  {\bibfield  {journal} {\bibinfo  {journal} {Appl. Phys. Lett.}\ }\textbf
  {\bibinfo {volume} {108}},\ \bibinfo {eid} {132402} (\bibinfo {year}
  {2016})}\BibitemShut {NoStop}%
\bibitem [{\citenamefont {Schliemann}(2016)}]{Schliemann2016}%
  \BibitemOpen
  \bibfield  {author} {\bibinfo {author} {\bibfnamefont {J.}~\bibnamefont
  {Schliemann}},\ }\href {https://arxiv.org/abs/1604.02026} {\  (\bibinfo
  {year} {2016})},\ \Eprint {http://arxiv.org/abs/1604.02026v1}
  {arXiv:1604.02026v1} \BibitemShut {NoStop}%
\bibitem [{\citenamefont {Zutic}\ \emph {et~al.}(2004)\citenamefont {Zutic},
  \citenamefont {Fabian},\ and\ \citenamefont {Sarma}}]{Zutic2004a}%
  \BibitemOpen
  \bibfield  {author} {\bibinfo {author} {\bibfnamefont {I.}~\bibnamefont
  {Zutic}}, \bibinfo {author} {\bibfnamefont {J.}~\bibnamefont {Fabian}}, \
  and\ \bibinfo {author} {\bibfnamefont {S.~D.}\ \bibnamefont {Sarma}},\ }\href
  {http://arxiv.org/abs/cond-mat/0405528} {\bibfield  {journal} {\bibinfo
  {journal} {Rev. Mod. Phys.}\ }\textbf {\bibinfo {volume} {76}},\ \bibinfo
  {pages} {323} (\bibinfo {year} {2004})}\BibitemShut {NoStop}%
\bibitem [{\citenamefont {D'yakonov}\ and\ \citenamefont
  {Kachorovskii}(1986)}]{Dyakonov1986}%
  \BibitemOpen
  \bibfield  {author} {\bibinfo {author} {\bibfnamefont {M.}~\bibnamefont
  {D'yakonov}}\ and\ \bibinfo {author} {\bibfnamefont {V.~Y.}\ \bibnamefont
  {Kachorovskii}},\ }\href@noop {} {\bibfield  {journal} {\bibinfo  {journal}
  {Sov. Phys. Semicond}\ }\textbf {\bibinfo {volume} {20}},\ \bibinfo {pages}
  {110} (\bibinfo {year} {1986})}\BibitemShut {NoStop}%
\bibitem [{\citenamefont {Iordanskii}\ \emph {et~al.}(1994)\citenamefont
  {Iordanskii}, \citenamefont {\mbox{Yu}. B.~Lyanda-Geller},\ and\
  \citenamefont {Pikus}}]{Iordanskii1994}%
  \BibitemOpen
  \bibfield  {author} {\bibinfo {author} {\bibfnamefont {S.~V.}\ \bibnamefont
  {Iordanskii}}, \bibinfo {author} {\bibnamefont {\mbox{Yu}.
  B.~Lyanda-Geller}}, \ and\ \bibinfo {author} {\bibfnamefont {G.~E.}\
  \bibnamefont {Pikus}},\ }\href
  {http://www.jetpletters.ac.ru/ps/1323/article\_20010.pdf} {\bibfield
  {journal} {\bibinfo  {journal} {JETP Lett.}\ }\textbf {\bibinfo {volume}
  {60}},\ \bibinfo {pages} {206} (\bibinfo {year} {1994})}\BibitemShut
  {NoStop}%
\bibitem [{\citenamefont {Kettemann}(2007)}]{Kettemann2007a}%
  \BibitemOpen
  \bibfield  {author} {\bibinfo {author} {\bibfnamefont {S.}~\bibnamefont
  {Kettemann}},\ }\href {http://link.aps.org/abstract/PRL/v98/e176808}
  {\bibfield  {journal} {\bibinfo  {journal} {Phys. Rev. Lett.}\ }\textbf
  {\bibinfo {volume} {98}},\ \bibinfo {pages} {176808} (\bibinfo {year}
  {2007})}\BibitemShut {NoStop}%
\bibitem [{\citenamefont {Mal'shukov}\ \emph {et~al.}(2005)\citenamefont
  {Mal'shukov}, \citenamefont {Wang}, \citenamefont {Chu},\ and\ \citenamefont
  {Chao}}]{Malshukov2005}%
  \BibitemOpen
  \bibfield  {author} {\bibinfo {author} {\bibfnamefont {A.~G.}\ \bibnamefont
  {Mal'shukov}}, \bibinfo {author} {\bibfnamefont {L.~Y.}\ \bibnamefont
  {Wang}}, \bibinfo {author} {\bibfnamefont {C.~S.}\ \bibnamefont {Chu}}, \
  and\ \bibinfo {author} {\bibfnamefont {K.~A.}\ \bibnamefont {Chao}},\ }\href
  {\doibase 10.1103/PhysRevLett.95.146601} {\bibfield  {journal} {\bibinfo
  {journal} {Phys. Rev. Lett.}\ }\textbf {\bibinfo {volume} {95}},\ \bibinfo
  {pages} {146601} (\bibinfo {year} {2005})}\BibitemShut {NoStop}%
\bibitem [{\citenamefont {Schwab}\ \emph {et~al.}(2006)\citenamefont {Schwab},
  \citenamefont {Dzierzawa}, \citenamefont {Gorini},\ and\ \citenamefont
  {Raimondi}}]{Schwab2006}%
  \BibitemOpen
  \bibfield  {author} {\bibinfo {author} {\bibfnamefont {P.}~\bibnamefont
  {Schwab}}, \bibinfo {author} {\bibfnamefont {M.}~\bibnamefont {Dzierzawa}},
  \bibinfo {author} {\bibfnamefont {C.}~\bibnamefont {Gorini}}, \ and\ \bibinfo
  {author} {\bibfnamefont {R.}~\bibnamefont {Raimondi}},\ }\href {\doibase
  10.1103/PhysRevB.74.155316} {\bibfield  {journal} {\bibinfo  {journal} {Phys.
  Rev. B}\ }\textbf {\bibinfo {volume} {74}},\ \bibinfo {pages} {155316}
  (\bibinfo {year} {2006})}\BibitemShut {NoStop}%
\bibitem [{\citenamefont {Wenk}\ and\ \citenamefont
  {Kettemann}(2010)}]{Wenk2010}%
  \BibitemOpen
  \bibfield  {author} {\bibinfo {author} {\bibfnamefont {P.}~\bibnamefont
  {Wenk}}\ and\ \bibinfo {author} {\bibfnamefont {S.}~\bibnamefont
  {Kettemann}},\ }\href {\doibase 10.1103/PhysRevB.81.125309} {\bibfield
  {journal} {\bibinfo  {journal} {Phys. Rev. B}\ }\textbf {\bibinfo {volume}
  {81}},\ \bibinfo {pages} {125309} (\bibinfo {year} {2010})}\BibitemShut
  {NoStop}%
\bibitem [{\citenamefont {Dyakonov}\ and\ \citenamefont
  {Perel}(1971)}]{Dyakonov1971}%
  \BibitemOpen
  \bibfield  {author} {\bibinfo {author} {\bibfnamefont {M.~I.}\ \bibnamefont
  {Dyakonov}}\ and\ \bibinfo {author} {\bibfnamefont {V.~I.}\ \bibnamefont
  {Perel}},\ }\href@noop {} {\bibfield  {journal} {\bibinfo  {journal} {Soviet
  Physics Jetp-Ussr}\ }\textbf {\bibinfo {volume} {33}},\ \bibinfo {pages}
  {1053} (\bibinfo {year} {1971})}\BibitemShut {NoStop}%
\bibitem [{\citenamefont {Pikus}\ and\ \citenamefont
  {Titkov}(1984)}]{pikustitkov}%
  \BibitemOpen
  \bibfield  {author} {\bibinfo {author} {\bibfnamefont {G.~E.}\ \bibnamefont
  {Pikus}}\ and\ \bibinfo {author} {\bibfnamefont {A.~N.}\ \bibnamefont
  {Titkov}},\ }in\ \href@noop {} {\emph {\bibinfo {booktitle} {Optical
  Orientation}}},\ \bibinfo {series} {Modern Problems in Condensed Matter
  Sciences}, Vol.~\bibinfo {volume} {8},\ \bibinfo {editor} {edited by\
  \bibinfo {editor} {\bibfnamefont {F.}~\bibnamefont {Meier}}\ and\ \bibinfo
  {editor} {\bibfnamefont {B.~P.}\ \bibnamefont {Zakharchenya}}}\ (\bibinfo
  {publisher} {North-Holland, Amsterdam},\ \bibinfo {year} {1984})\
  Chap.~\bibinfo {chapter} {3}, p.\ \bibinfo {pages} {73�131}\BibitemShut
  {NoStop}%
\bibitem [{\citenamefont {Hikami}\ \emph {et~al.}(1980)\citenamefont {Hikami},
  \citenamefont {Larkin},\ and\ \citenamefont {Nagaoka}}]{nagaoka}%
  \BibitemOpen
  \bibfield  {author} {\bibinfo {author} {\bibfnamefont {S.}~\bibnamefont
  {Hikami}}, \bibinfo {author} {\bibfnamefont {A.~I.}\ \bibnamefont {Larkin}},
  \ and\ \bibinfo {author} {\bibfnamefont {Y.}~\bibnamefont {Nagaoka}},\ }\href
  {\doibase 10.1143/PTP.63.707} {\bibfield  {journal} {\bibinfo  {journal}
  {Prog. Theor. Phys.}\ }\textbf {\bibinfo {volume} {63}},\ \bibinfo {pages}
  {707} (\bibinfo {year} {1980})}\BibitemShut {NoStop}%
\bibitem [{\citenamefont {Knap}\ \emph {et~al.}(1996)\citenamefont {Knap},
  \citenamefont {Skierbiszewski}, \citenamefont {Zduniak}, \citenamefont
  {Litwin-Staszewska}, \citenamefont {Bertho}, \citenamefont {Kobbi},
  \citenamefont {Robert}, \citenamefont {Pikus}, \citenamefont {Pikus},
  \citenamefont {Iordanskii}, \citenamefont {Mosser}, \citenamefont
  {Zekentes},\ and\ \citenamefont {\mbox{Yu}. B.~Lyanda-Geller}}]{Knap1996}%
  \BibitemOpen
  \bibfield  {author} {\bibinfo {author} {\bibfnamefont {W.}~\bibnamefont
  {Knap}}, \bibinfo {author} {\bibfnamefont {C.}~\bibnamefont
  {Skierbiszewski}}, \bibinfo {author} {\bibfnamefont {A.}~\bibnamefont
  {Zduniak}}, \bibinfo {author} {\bibfnamefont {E.}~\bibnamefont
  {Litwin-Staszewska}}, \bibinfo {author} {\bibfnamefont {D.}~\bibnamefont
  {Bertho}}, \bibinfo {author} {\bibfnamefont {F.}~\bibnamefont {Kobbi}},
  \bibinfo {author} {\bibfnamefont {J.~L.}\ \bibnamefont {Robert}}, \bibinfo
  {author} {\bibfnamefont {G.~E.}\ \bibnamefont {Pikus}}, \bibinfo {author}
  {\bibfnamefont {F.~G.}\ \bibnamefont {Pikus}}, \bibinfo {author}
  {\bibfnamefont {S.~V.}\ \bibnamefont {Iordanskii}}, \bibinfo {author}
  {\bibfnamefont {V.}~\bibnamefont {Mosser}}, \bibinfo {author} {\bibfnamefont
  {K.}~\bibnamefont {Zekentes}}, \ and\ \bibinfo {author} {\bibnamefont
  {\mbox{Yu}. B.~Lyanda-Geller}},\ }\href
  {http://link.aps.org/doi/10.1103/PhysRevB.53.3912} {\bibfield  {journal}
  {\bibinfo  {journal} {Phys. Rev. B}\ }\textbf {\bibinfo {volume} {53}},\
  \bibinfo {pages} {3912} (\bibinfo {year} {1996})}\BibitemShut {NoStop}%
\bibitem [{\citenamefont {Kammermeier}\ \emph {et~al.}(2016)\citenamefont
  {Kammermeier}, \citenamefont {Wenk}, \citenamefont {Schliemann},
  \citenamefont {Heedt},\ and\ \citenamefont {Sch\"apers}}]{Kammermeier2016}%
  \BibitemOpen
  \bibfield  {author} {\bibinfo {author} {\bibfnamefont {M.}~\bibnamefont
  {Kammermeier}}, \bibinfo {author} {\bibfnamefont {P.}~\bibnamefont {Wenk}},
  \bibinfo {author} {\bibfnamefont {J.}~\bibnamefont {Schliemann}}, \bibinfo
  {author} {\bibfnamefont {S.}~\bibnamefont {Heedt}}, \ and\ \bibinfo {author}
  {\bibfnamefont {T.}~\bibnamefont {Sch\"apers}},\ }\href {\doibase
  10.1103/PhysRevB.93.205306} {\bibfield  {journal} {\bibinfo  {journal} {Phys.
  Rev. B}\ }\textbf {\bibinfo {volume} {93}},\ \bibinfo {pages} {205306}
  (\bibinfo {year} {2016})}\BibitemShut {NoStop}%
\bibitem [{\citenamefont {Beenakker}\ and\ \citenamefont {van
  Houten}(1991)}]{Beenakker1991}%
  \BibitemOpen
  \bibfield  {author} {\bibinfo {author} {\bibfnamefont {C.}~\bibnamefont
  {Beenakker}}\ and\ \bibinfo {author} {\bibfnamefont {H.}~\bibnamefont {van
  Houten}},\ }in\ \href {\doibase
  http://dx.doi.org/10.1016/S0081-1947(08)60091-0} {\emph {\bibinfo {booktitle}
  {Semiconductor Heterostructures and Nanostructures}}},\ \bibinfo {series}
  {Solid State Physics}, Vol.~\bibinfo {volume} {44},\ \bibinfo {editor}
  {edited by\ \bibinfo {editor} {\bibfnamefont {H.}~\bibnamefont {Ehrenreich}}\
  and\ \bibinfo {editor} {\bibfnamefont {D.}~\bibnamefont {Turnbull}}}\
  (\bibinfo  {publisher} {Academic Press},\ \bibinfo {year} {1991})\ pp.\
  \bibinfo {pages} {1 -- 228}\BibitemShut {NoStop}%
\bibitem [{\citenamefont {Malshukov}\ \emph {et~al.}(1997)\citenamefont
  {Malshukov}, \citenamefont {Chao},\ and\ \citenamefont
  {Willander}}]{Malshukov1997}%
  \BibitemOpen
  \bibfield  {author} {\bibinfo {author} {\bibfnamefont {A.~G.}\ \bibnamefont
  {Malshukov}}, \bibinfo {author} {\bibfnamefont {K.~A.}\ \bibnamefont {Chao}},
  \ and\ \bibinfo {author} {\bibfnamefont {M.}~\bibnamefont {Willander}},\
  }\href {http://link.aps.org/doi/10.1103/PhysRevB.56.6436} {\bibfield
  {journal} {\bibinfo  {journal} {Phys. Rev. B}\ }\textbf {\bibinfo {volume}
  {56}},\ \bibinfo {pages} {6436} (\bibinfo {year} {1997})}\BibitemShut
  {NoStop}%
\bibitem [{\citenamefont {Ganichev}\ \emph {et~al.}(2002)\citenamefont
  {Ganichev}, \citenamefont {Danilov}, \citenamefont {Bel'kov}, \citenamefont
  {Ivchenko}, \citenamefont {Bichler}, \citenamefont {Wegscheider},
  \citenamefont {Weiss},\ and\ \citenamefont {Prettl}}]{Ganichev2002}%
  \BibitemOpen
  \bibfield  {author} {\bibinfo {author} {\bibfnamefont {S.~D.}\ \bibnamefont
  {Ganichev}}, \bibinfo {author} {\bibfnamefont {S.~N.}\ \bibnamefont
  {Danilov}}, \bibinfo {author} {\bibfnamefont {V.~V.}\ \bibnamefont
  {Bel'kov}}, \bibinfo {author} {\bibfnamefont {E.~L.}\ \bibnamefont
  {Ivchenko}}, \bibinfo {author} {\bibfnamefont {M.}~\bibnamefont {Bichler}},
  \bibinfo {author} {\bibfnamefont {W.}~\bibnamefont {Wegscheider}}, \bibinfo
  {author} {\bibfnamefont {D.}~\bibnamefont {Weiss}}, \ and\ \bibinfo {author}
  {\bibfnamefont {W.}~\bibnamefont {Prettl}},\ }\href {\doibase
  10.1103/PhysRevLett.88.057401} {\bibfield  {journal} {\bibinfo  {journal}
  {Phys. Rev. Lett.}\ }\textbf {\bibinfo {volume} {88}},\ \bibinfo {pages}
  {057401} (\bibinfo {year} {2002})}\BibitemShut {NoStop}%
\end{thebibliography}%
\end{document}